\documentstyle[aps,psfig,epsfig]{revtex}

\newcommand{\ep}{\epsilon}

\newcommand{\pa}{\partial}
\newcommand{\td}{\tilde}
\newcommand{\kap}{\kappa}
\newcommand{\lrightarrow}{-\!\!-\!\!\!\longrightarrow}
\begin{document}

\draft
\title{Static Solution in Source-Free SU(2) Yang-Mills Theory}   
\author{Xiao-Jun Wang\footnote{E-mail address: wangxj@mail.ustc.edu.cn}}
\address{Center for Fundamental Physics,
University of Science and Technology of China\\
Hefei, Anhui 230026, P.R. China}
\author{Mu-Lin Yan\footnote{E-mail address: mlyan@staff.ustc.edu.cn}}
\address{CCST(World Lad), P.O. Box 8730, Beijing, 100080, P.R. China \\
  and\\
 Center for Fundamental Physics,
University of Science and Technology of China\\
Hefei, Anhui 230026, P.R. China\footnote{mail
address}}
\date{\today}
\maketitle

\begin{abstract}
We show that a non-trivial topological effect breaks the conformal
invariance of pure Yang-Mills theory. Thus it is possible that classic
particle-like solutions exists in pure Yang-Mills theory.
We find a static, non-singular solution in source-free SU(2) Yang-Mills
theory in four-dimensional Minkowski space. This solution is a stable
soliton characterized by non-trivial topology and imaginary $A_0^a$, i.e.,
$A_0^aA_0^a<0$. It yields hermitian Hamilton, and finite, positive energy.
\end{abstract}

\section*{}
The purpose of this paper is to study an important aspect of 
Yang-Mills theory: Does there exist classic, static,
non-sigular solution when non-trivial topology is present? Taditionally,
the pure Yang-Mills theory does not admit classical particle-like
solutions\cite{Coleman75,Deser76,Pagel77,Coleman77}. More precisely, this
famous result asserts that there exist no finite-energy non-singular
solutions to the four-dimensional static Yang-Mills
equations\cite{Coleman77}. Non-existence of static solutions can be
related to conformal invariance of the Yang-Mills theory, which implies
that the stress-energy tensor is traceless:
$T_{\mu}^{\mu}=0=-T_{00}+T_{ii}$, where $\mu=0,...,3,\;i=1,2,3,$ and
Minkowski metric is taken $\eta_{\mu\nu}={\rm diag}\{1,-1,-1,-1\}$. The
positivity of energy density $T_{00}$ requires that the sum of the
principal pressures $T_{ii}$ is everywhere positive, i.e., the Yang-Mills
matter is repulsive. 

However, in process to obtain the above result, all surface terms are
droped. In the other words, $T_\mu^\mu=0$ is only a result in the case of
topological trivial. In static case, if a non-trivial topological effect
exists here:
\begin{eqnarray}\label{1}
\int d^3\vec{x}\pa_i(\bar{F}_{i0}^aA_\nu^a)={\rm constant},  
\end{eqnarray}
the conformal
invariance of pure Yang-Mills theory is obviously broken, where
$\bar{F}_{\mu\nu}^a(\vec{x})=\pa_\mu A_\nu^a-\pa_\nu A_\mu^a
2g\ep^{abc}A_\mu^bA_\nu^c,\;(a=1,...,N^2-1)$ are SU(N) Yang-Mills field 
strength. In this present paper, the ``static'' means that all gauge
fields and gauge transformations are time-indepenedent. The broken of
conformal invariance can be confirmed simply by a scale transformation
$\vec{x}\rightarrow\vec{x}/\lambda$: The right side of eq.~(\ref{1})
implies that the left side of eq.~(\ref{1}) should be invariant
at $|x|\rightarrow \infty$. It conflicts with the conformal invariance of
pure Yang-Mills theory, which requires that left side of eq.~(\ref{1})
should be proportional to $\lambda$ or vanish at $|x|\rightarrow \infty$.

Let us further clarify what role this non-trivial topological effect
plays. Noether's theorem yields conservation stress-energy tensor as
follow
\begin{eqnarray}\label{2}
T_{\mu\nu}=\frac{1}{4}\eta_{\mu\nu}F_{\alpha\beta}^aF^{\alpha\beta a} 
       -F_{\mu\beta}^a\pa_\nu A^{\beta a}.
\end{eqnarray}     
When topology is trivial and Yang-Mills field is no-singular, a
surface term can be freely added in stress-energy tensor for obtaining
gauge invariant form,
\begin{eqnarray}\label{3}
T_{\mu\nu}\rightarrow T_{\mu\nu}+\pa^\lambda(F_{\mu\lambda}^aA_\nu^a)
=\frac{1}{4}\eta_{\mu\nu}F_{\alpha\beta}^aF^{\alpha\beta a}
 -F_{\mu\beta}^aF^{\mu\beta a}.
\end{eqnarray} 
Eq.~(\ref{3}) leads to $T_\mu^\mu=0$ in four-dimensional spacetime.
However, for the case of non-trivial topology presence(eq.~(\ref{1})),
this surface term can not be added, since it obviously changes physical
energy-momentum. So that $T_\mu^\mu=0$ can not be yielded in this case. 

When Yang-Mills fields are static, the Coleman-Deser
equation\cite{Coleman75,Deser76,Wu76} 
\begin{eqnarray}\label{4}
2\int d^3\vec{x}T_i^i=
\int d^3\vec{x}\{\frac{1}{2}\bar{F}_{ij}^a\bar{F}_{ij}^a
   +\bar{F}_{i0}^a\bar{F}_{i0}^a\}=0
\end{eqnarray}
can be obtained even though the non-trivial topology is present, since we
need only static Yang-Mills equations and requirement of finite energy to
obtain it . This equation provides some constrains for static solution of
Yang-Mills equations. For trivial topology, the term
$\bar{F}_{i0}^a\bar{F}_{i0}^a$ vanishes due to static Yang-Mills
equations. It leads to the trivial solution
$\bar{F}_{i0}^a=\bar{F}_{ij}^a=0$. However, for non-trivial topology,
$\bar{F}_{i0}^a\bar{F}_{i0}^a$ does not vanish. Then Coleman-Deser
equation admits static solution of pure Yang-Mills theory when
$\bar{F}_{ij}^a\bar{F}_{ij}^a<0$ or $\bar{F}_{i0}^a\bar{F}_{i0}^a<0$. 
Meanwhile, the static Hamilton density
\begin{eqnarray}\label{5}
{\cal H}=\frac{1}{4}\bar{F}_{ij}^a\bar{F}_{ij}^a
   -\frac{1}{2}\bar{F}_{i0}^a\bar{F}_{i0}^a
\end{eqnarray}
implies that only $\bar{F}_{i0}^a\bar{F}_{i0}^a<0$ is allowed. It
indicates that the static solution exists only for imaginary $A_0^a$. 
It means that $A_0^a$ have to be continued to complex plane analytically.
The complex solutions were studied by Dolan\cite{Dolan77} in Euclidian
space. He obtained some Abelian-like solutions with zero action. In this
present paper, we will actually study imaginary solution of $A_0$ with in
Minkowski space. It will yield finite, no-zero energy. In the static pure
Yang-Mills theory, at least, this imaginary $A_0^a$ is allowed by all
fundamental principle, such as positvity of energy, hermitian of Hamilton,
etc..

In the following we will try to find an analytic, static, non-singular
soliton soultion of source-free SU(2) Yang-Mills equations. We can define
``static dual'' in Minkowski space,
\begin{eqnarray}\label{6}
\td{\bar{F}}_{ij}^a=i\ep_{ijk}\bar{F}_{k0}^a,\hspace{1.5in}
\td{\bar{F}}_{k0}^a=-\frac{i}{2}\ep_{ijk}\bar{F}_{ij}^a.
\end{eqnarray}
where $a=1,2,3$. The ``static dual'' in the above equation satisfy
\begin{eqnarray}\label{7}
\td{\td{\bar{F}}}_{ij}^{\atop {\scriptstyle a}}
=\bar{F}_{ij}^a,\hspace{1.5in}
\td{\td{\bar{F}}}_{i0}^{\atop {\scriptstyle a}}=\bar{F}_{i0}^a.
\end{eqnarray}
It is easily to check that, the static Yang-Mills equations and
Coleman-Deser equation~(\ref{4}) will be
automatically statisfied if field strengths are ``static self-dual''
\begin{eqnarray}\label{8}
\bar{F}_{\mu\nu}^a=\td{\bar{F}}_{\mu\nu}^a.
\end{eqnarray}
In addition, the ``static self-dual'' also yielded minimum energy of the
system. 
\footnote{The solutions of eq.~(\ref{8}) may be nameed as ``static 
instanton''. However, it must be pointed out that the usual static
instanton is not well-defined: The instanton is related to tunneling
effects of quantum mechanics which exsits only in Euclidean space
(imaginary time). In the case absence time, there exists no tunneling
effects, so that solutions of eq~(\ref{8}) is different from traditional
instanton.}

For obtaining an explicit analytic solution, in this paper we take
a spherical sysmmetry ansatz for gauge fields
\begin{eqnarray}\label{9}
\left\{ {{\displaystyle A_{ia}=\frac{1}{g}\frac{f(r)}{r}\ep_{ian}
\hat{x}_n,}\atop {\displaystyle
\!\!\!\!A_{0a}=\frac{i}{g}\phi(r)\hat{x}^a,}}\right.\hspace{1.5in}
\hat{x}={x_a\over r}.
\end{eqnarray}
Than eqs.~(\ref{8}) and (\ref{9}) lead to
\begin{eqnarray}\label{10}
\left\{ {{\displaystyle f'=-\phi(1-2f),}\atop {\displaystyle
\phi'=-\frac{2}{r^2}f(1-f),}}\right.
\end{eqnarray}
with $f'=\frac{d}{dr}f,\;\phi'=\frac{d}{dr}\phi$. To obtain the above 
equation, the identity,
\begin{eqnarray}\label{other}
\ep_{jan}\hat{x}_i\hat{x}_n-\ep_{ian}\hat{x}_j\hat{x}_n
+\ep_{ijn}\hat{x}_a\hat{x}_n=\ep_{ija},
\end{eqnarray}
has been used. The eq.~(\ref{10}) can reduce to static Liouville equation
\cite{Jackiw82}. Its solutions are well-know. An analytic, non-sigular
solution is, 
\begin{eqnarray}\label{11}
\left\{ {{\displaystyle f=\frac{1}{2}(1-\frac{\kap r}{\sinh(\kap r)}),}
\atop {\displaystyle \phi=\frac{1}{2r}(1-\frac{\kap r}{\tanh(\kap r)}),}} 
\right.
\end{eqnarray}
where $\kap$ is a positive integral constant with mass-dimension. The
solutions~(\ref{11}) lead to their asymptotic behaviour as follows
\begin{eqnarray}\label{12}
&&f\stackrel{r\rightarrow 0}\lrightarrow\frac{1}{12}\kap^2 r^2,
\hspace{1.12in}\phi\stackrel{r\rightarrow 0}\lrightarrow
 -\frac{1}{6}\kap^2 r,\nonumber\\
&&f\stackrel{r\rightarrow \infty}\lrightarrow{1\over 2}-e^{-\kap r},
\hspace{1in}\phi\stackrel{r\rightarrow \infty}\lrightarrow
-\frac{\kap}{2}+\frac{1}{2r}.
\end{eqnarray}
Thus it is suprised that $A_0\rightarrow$ constant instead of zero at
$r\rightarrow\infty$ (it is different from instanton and monopole), but
while all $F_{\mu\nu}$ still fall off as $r^{-2}$.. 

Due to ``static self-dual'' of field strength, the topological mass
(energy) of the soliton is
\begin{eqnarray}\label{13}
M&=&\int d^3\vec{x}{\cal H}=-\int d^3\vec{x}
 \bar{F}_{i0}^a\bar{F}_{i0}^a=-\int d^3\vec{x}
 \pa_i(A_0^a\bar{F}_{i0}^a)\nonumber\\
&=&-\int d^3\vec{x}\pa_i(A_0^a\pa_iA_{0}^a)=\frac{\pi}{g^2}\kap.
\end{eqnarray}
Therefore, the integral constant $\kap$ can be interpreted as an unit of
energy. Moreover, we can see that the topological effect in our solution
is characterized by the topological mass, which is determined by asympotic
behaviour of $A_0^a$ at $r\rightarrow\infty$. The energy density
distribution function is
\begin{eqnarray}\label{14}
\rho(r)=-\bar{F}_{i0}^a\bar{F}_{i0}^a=g^{-2}\{\phi\phi''+
\frac{2}{r}\phi\phi'+\phi'^2\}.
\end{eqnarray}
The energy density distribution is shown in fig. 1. Effective radius of
the soliton (soliton size) is usually defined as half maximum width of
spectral distribution. Then From the fig. 1 we can see that the effective
radius of this soliton is $r_0=1.3\kap^{-1}$. It is interesting that
$r_0M$ is $\kap$-independent. The fig. 1 also shows that this soliton is
stable for any boundary conditions, i.e., fixed $\kap$.

We have shown that there exist static topological solution with finite
energy in pure Yang-Mills theory. This fact indicates that the
massless gauge particles in Yang-Mills theory can become
static, massive particles due to non-linear self-interaction. It is
interesting to compare with U(1) electromagnetic theory, in which the
photon can not be static and massive forever. In our solution, the zero
component of Yang-Mills field is imaginary. However, it does not cause any
problem for static pure Yang-Mills theory. In fact, the ansatz~(\ref{9})
naturally satisfies the Coulomb gauge condition, $\pa_i A_i=0$.
Then for the static case, Lorentz condition $\pa_\mu A^\mu=0$ allows $A_0$
to be arbitrary time-independent function (no matter what it is real or
imaginary).

The confinement of gauge particles is an active subject (specially in
quantum chromodynamics). 
Although the solution~(\ref{11}) yields finite energy, the confinement of
gauge particles is still allowed here. There are two different mechanisms
to achieve the confinement: The first one is to take a limit of 
solution~(\ref{11}), i.e., $\kap\rightarrow\infty$. It is obvious that, in
this limit topological mass $M\rightarrow\infty$ and effective radius
$r_0\rightarrow 0$. It means that the static point-like particles will
generate infinite energy. The second one is that there exist some
other singular solutions for eq.~(\ref{11}). For example, the
solution
\begin{eqnarray}\label{15}
f=r\phi=\frac{\lambda}{r+2\lambda}
\end{eqnarray}     
leads gauge fields to be singular at $r=0$. Thus energy obtained from
solution~(\ref{15}) is infinite. These two mechanisms can be
distinguished by asymptotic behaviour of $\phi$ (or $A_0$) at 
$r\rightarrow\infty$. In the first one, $\phi\rightarrow\infty$ and
$\phi'\rightarrow 0$ at $r\rightarrow\infty$ when $\kap\rightarrow\infty$.
Meanwhile, in the second one, $\phi\rightarrow r^{-2}\rightarrow 0$ at
$r\rightarrow\infty$. The above discussion indicates that both of
confinement and no-confinement are allowed in pure Yang-Mills
theory. Whether or not presence of confinement is determined by asymptotic
behaviour of $A_0^a$ at $r\rightarrow\infty$.     

\begin{figure}[htpb]
   \centerline{\psfig{figure=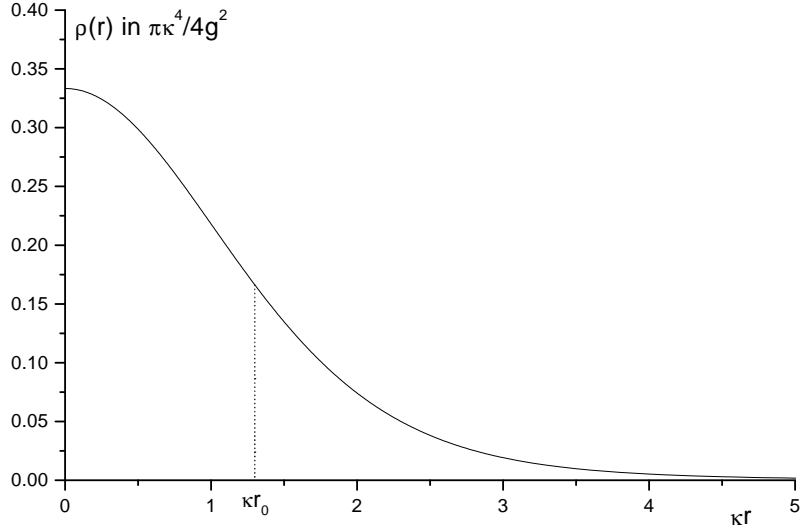,width=5in}}
 \centering
\begin{minipage}{5in}
   \caption{The density distribution of topological soliton. It implies 
that the effective radius(width of wave packet) is $r_0=1.3\kap^{-1}$.}
\end{minipage}
\end{figure}
  
To conclude, we obtain a static spherical sysmmetry no-singular soliton
solution in source-free Yang-Mills theory when a non-trivial
topology is present. This topological soliton is stable, and has finite,
positive and minimum mass (energy). Value of the topological mass can be
fixed by effective radius of the solution. At the limit of point-particle,
the gauge fields will be confinement (with infinite energy). The
topological effect is determined by asymptotic behaviour of $A_0^a$ at
$r\rightarrow\infty$. 

\begin{center}
{\bf ACKNOWLEDGMENTS}
\end{center}
The authors acknowledge R. Jackiw for serveral valuable comments. They
also thank B.-Q. Ma for valuable discussion. This work is partially
supported by NSF of China through C. N. Yang and the Grant LWTZ-1298 of
Chinese Academy of Science. The authors also acknowledges R. Jackiw for
serveral valuable comments

\end{document}